\documentclass{pasj00}
\draft

\RelativeFigurePath{.}

\begin{document}
\SetRunningHead{Author(s) in page-head}{Running Head}

\title{Optical Dual-Band Photometry and Spectroscopy of the WZ Sge-Type Dwarf Nova EZ Lyn during the 2010 Superoutburst}

\author{%
   Mizuki \textsc{Isogai}\altaffilmark{1,2},
   Akira \textsc{Arai}\altaffilmark{1,3},
   Atsunori \textsc{Yonehara}\altaffilmark{1},
   Hideyo \textsc{Kawakita}\altaffilmark{1},
   Makoto \textsc{Uemura}\altaffilmark{4},
   \and
   Daisaku \textsc{Nogami}\altaffilmark{5,6}}

\altaffiltext{1}{Koyama Astronomical Observatory, Kyoto Sangyo University, 
         Motoyama, Kamigamo, Kita-Ku, Kyoto, Kyoto 603-8555, Japan}
\altaffiltext{2}{National Astronomical Observatory of Japan, 
         2-21-1 Osawa, Mitaka, Tokyo 181-8588, Japan}
\email{mizuki.isogai@nao.ac.jp}
\altaffiltext{3}{Nishi-Harima Astronomical Observatory, Center for Astronomy, University of Hyogo, 407-2 Nishigaichi, Sayo, Hyogo 679-5313, Japan}
\altaffiltext{4}{Hiroshima Astrophysical Science Center, Hiroshima University, Kagamiyama 1-3-1, Higashi-Hiroshima, Hiroshima 739-8526, Japan}
\altaffiltext{5}{Kwasan and Hida Observatories, Kyoto University, Yamashina-ku, Kyoto, Kyoto 607-8471, Japan}
\altaffiltext{6}{Department of Astronomy, Kyoto University, Kitashirakawa Oiwake-cho, Sakyo-ku, Kyoto, Kyoto 606-8502, Japan}


\maketitle

\begin{abstract}
We performed optical simultaneous dual-band (SDSS $g'$- and $i'$- band) photometry and low-resolution spectroscopy for the WZ Sge-type dwarf nova EZ Lyn during its 2010 superoutburst.
Dual-band photometry revealed that the $g'-i'$ color reddened with a decrease in brightness, during the main superoutburst and the following rebrightening phase, whereas the color became bluer with a further decrease in brightness during the slow, final decline phase.
With a fit to our photometric results by a blackbody function, we estimated the disk radius ratio (ratio of the disk radius to the binary separation) and compared this with that of V455 And, a WZ Sge-type object that did not show any rebrightening in the 2007 superoutburst. 
The comparison revealed: (1) the disk radius ratio of EZ Lyn decreased more slowly than that of V455 And; and (2) the radius ratio of EZ Lyn at the end of the main superoutburst was larger than that of the V455 And.
These results favor the mass reservoir model for the mechanism of rebrightening.
%
During both the superoutburst plateau and subsequent rebrightening phase, H$\alpha$ and H$\beta$ lines were detected.
The H$\alpha$ line showed a double-peak profile from which we estimated the disk radius ratio.
The comparison of this ratio with that derived by photometry, indicates that the H$\alpha$ disk was larger than the photometric one, which suggests that the optically thin gas was extended to the outer region more than the optically thick gas disk and was possibly responsible for the rebrightening phenomenon.
%
Time-series dual-band photometry during the main superoutburst revealed that color variations during the early superhump show roughly the same behavior as that of V455 And, whereas color variations during the ordinary superhump display clear anticorrelation with brightness, in contrast to that seen in the V455 And.
Here, we discuss different color behaviors.
\end{abstract}

\section{Introduction}
Dwarf novae are a subclass of cataclysmic variables with a type of close binary system consisting of a white dwarf and a late-type main-sequence (secondary) star.
The matter lost from the secondary star filling its Roche-Lobe is transferred to the white dwarf via the Lagrangian point L1, and makes an accretion disk around the white dwarf.
One of the most peculiar events in dwarf novae are its sudden increases in brightness, described as (normal) outbursts.
It is currently understood that these outbursts are triggered by thermal instability in the accretion disk (e.g. \cite{hoshi1979}).

SU Ursae Major (SU UMa)-type dwarf novae are a subgroup of dwarf novae.
These objects show two types of outburst: a normal outburst and a {\it superoutburst}.
The superoutburst has a larger amplitude, by 0.2--0.4 mag, and a longer duration (about 10 days) than the normal outburst.
During the superoutburst, the object shows short-term periodic modulations, called {\it superhumps}, with an amplitude of about 0.3--0.4 mag, and with a period a few percent greater than the orbital period \citep{warner1985}.
It is widely accepted that this superoutburst is caused by the tidal instability of the disk (\cite{whitehurst1988}; \cite{osaki1989}).
The model describes that when a disk expands beyond the 3:1 resonance radius during a normal outburst, a tidal instability is excited.
The disk is then deformed to an eccentric form by a strong tidal torque, and undergoes a slow precession (\cite{vogt1982}; \cite{osaki1985}; \cite{whitehurst+king1991}).
The more effective tidal dissipation, caused by the deformed elliptical disk, leads to a bright superoutburst.

WZ Sagittae (WZ Sge)-type dwarf novae are an extreme subgroup of SU UMa-type dwarf novae.
The peculiarities of these objects are: (1) no normal outburst; (2) a long superoutburst recurrence interval (several years up to decades); and (3) a larger superoutburst amplitude ($\geq 6$ mag) than in SU UMa-type objects.
In addition, many (but not all) WZ Sge-type objects show two unique behaviors: `early superhumps' observed only in the very early phase of the superoutburst; and post-superoutburst rebrightening \citep{kato+2001}.

The early superhump is a modulation of brightness showing two peaks during a period that is almost the same as the orbital period (see \cite{kato+1996}; \cite{ishioka+2002}; \cite{osaki+meyer2002}; \cite{patterson+2002}).
To explain this modulation, \citet{osaki+meyer2002} proposed a model of a two-armed spiral structure excited by tidal dissipation produced by a 2:1 resonance.
Recently, simultaneous multicolor photometry of WZ Sge-type objects revealed that the color becomes red at the maximum of the early superhump (V455 And: \cite{matsui+2009}; OT J012059.6+325545: \cite{nakagawa+2013}).
\citet{uemura+2012} published a model code to reconstruct the disk height structure from multicolor data on the assumption that early superhumps are caused by non-uniformity of disk height structure.
The application of this code to both objects succeeded in reproducing the arm-like pattern in the reconstructed disk height map (see \cite{uemura+2012} for the case of V455 And, and \cite{nakagawa+2013} for OT J012059.6+325545).

Rebrightening is a phenomenon where brightness increases again after the end of the main superoutburst.
This phenomenon is seen in WZ Sge-type and some SU UMa-type objects \citep{kato+2004}, and is classified by light curve into three types: (1) single short type; (2) short, repetitive rebrightening; and (3) long-lived plateau type \citep{imada+2006}.
As a mechanism for this rebrightening phenomenon, two competing models have been proposed.
One model is the {\it `mass reservoir model'} where rebrightening is caused by delayed accretion of gas left in the outermost region, expanded beyond the 3:1 resonance radius by the superoutburst (\cite{kato+1998}; \cite{hellier2001}; \cite{osaki+meyer2001}).
The other model is the {\it `enhanced mass transfer model'} where rebrightening is caused by enhanced mass transfer from the secondary star irradiated by the accretion disk and central white dwarf, brightened by the superoutburst (\cite{patterson+1998}; \cite{hameury+2000}).
These models predict a different temporal evolution of the disk radius; the former predicts a {\it slow} decrease, whereas the latter model predicts a {\it rapid} decrease. 
Multicolor observations that provide information on the size of the emitting region can possibly evaluate these models.

In this paper, we report the results of our photometric and spectroscopic observations of WZ Sge-type dwarf nova EZ Lyncis (EZ Lyn) (also known as SDSS J080434.20+510349.1) during its 2010 superoutburst.
From daily variations in brightness, color, and H$\alpha$ line profile, we estimated the disk radius ratio (ratio of the disk radius to the binary separation), compared it with that of the V455 Andromedae (V455 And, a WZ Sge-type object that did not show any rebrightening in the 2007 superoutburst), and discussed the difference between the two objects.
From the short-term variations in brightness and color in a day, we calculated the variation of the emitting size and discussed the superhump source.

EZ Lyn was discovered by \citet{szkody+2006}.
\citet{pavlenko+2007} found this object was in superoutburst in 2006.
Although a lack of observation prevented detection of the early superhump, it was classified as a WZ Sge-type dwarf nova on the basis of the estimated amplitude of the superoutburst (about 7 mag) and rebrightening, which were detected 11 times (\cite{pavlenko+2007}; \cite{kato+2009b}).
In 2010, this object caused a second superoutburst only 4 years after the first.
In this superoutburst, EZ Lyn showed rebrightening on six occasions.
The periods of early and ordinary superhumps were studied by \citet{kato+2012}.
Among WZ Sge-type stars, EZ Lyn has several peculiarities: a short recurrence time; brightenings and mini-outbursts during quiescence (\cite{szkody+2006}; \cite{zharikov+2008}); and non-radial pulsations seen only during part of the post-outburst phase (\cite{pavlenko+2007}; \cite{pavlenko+2009}; \cite{pav+mal2009}; \cite{pavlenko+2012}; and \cite{szkody+2013}).
When the spectral energy distribution (SED) model was fit with observations in quiescence, the mass and surface temperature of the white dwarf in EZ Lyn were estimated at $M_\mathrm{WD} \ge 0.7\MO$ and $T \sim 12000K$ (\cite{zharikov+2013}).

In section 2, the observations and data reduction are described.
In section 3, our observational results are displayed.
We discuss our results in section 4, and summarize our study in section 5.

\section{Observation}

EZ Lyn was observed both photometrically and spectroscopically.
All of the observations were carried out using the Araki Telescope (1.3-m diameter; F/10) at Koyama Astronomical Observatory at Kyoto-Sangyo University, Japan.

\subsection{Photometry}
The photometry of EZ Lyn was performed between September 18, 2010 and January 7, 2011  using a dual-band imager, ADLER (Araki telescope DuaL-band imagER).

The ADLER has two dichroic mirrors whose transition wavelengths are 480\,nm and 670\,nm, respectively.
The dichroic mirror splits F/10, converging light from the telescope into two beams.
Each beam is converged to F/6 by the F-conversion lens set, and is focused on each charge-coupled device (CCD) camera after passing through two filter wheels with six holes. 
The CCD cameras are water-cooled, Spectral Instruments (850 series), capable of cooling to 183$K$ with 293$K$ circulated water.
These cameras have e2v CCD 42-40 (2048 $\times$ 2048 pixels, 13.5\,$\micron$/pixel corresponding to 0.357 arcsec/pixel, 12 arcmin square for field of view), back and front illuminated chips adopted by short and long wavelength cameras, respectively.

\begin{figure}
  \begin{center}
    \FigureFile(80mm,120mm){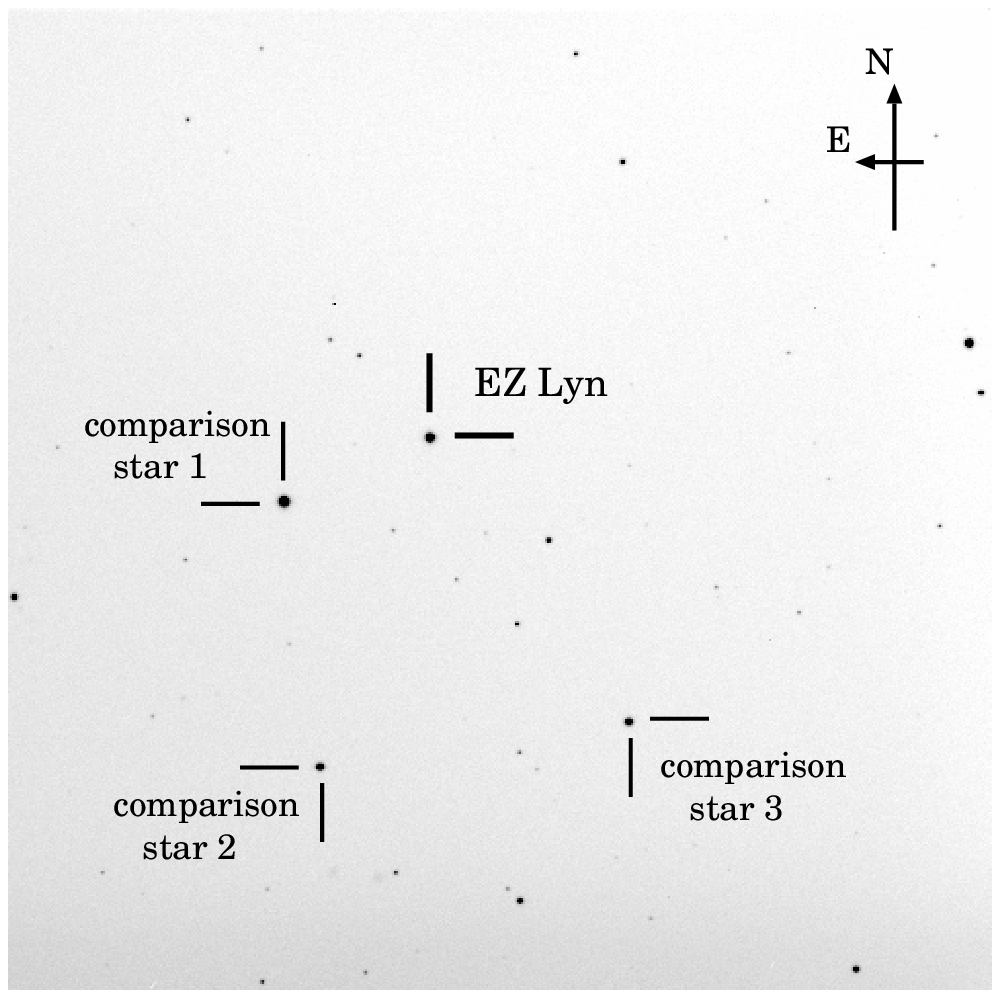}
  \end{center}
  \caption{SDSS $g'$-band CCD image of the field of EZ Lyn. This image was obtained by the ADLER, on September 18, 2010. The field of view is $12' \times 12'$. EZ Lyn and comparison stars are identified using black bars.}\label{fig:comparison-star-position}
\end{figure}

EZ Lyn was observed with a 670\,nm-split dichroic mirror, and the Sloan Digital Sky Survey (SDSS) $g'$ and $i'$ filters.
The seeing condition (full width at half maximum (FWHM)) was 2.0--5.3 and 1.7--4.9 arcsec for $g'$ and $i'$ bands, respectively.
%
The observational data were reduced using the image reduction and analysis facility (IRAF)  with the standard techniques of photometry (dark-subtraction, flat-fielding, and aperture photometry with noao.digiphot.daophot).
We adopted three comparison stars: USNO-B1 1410-0196618 (C1) for almost all observations, USNO-B1 1409-0197659 (C2) for November 1, 2010 and December 1, 2010, and USNO-B1 1410-0196577 (C3) for January 7, 2011.
The positions of EZ Lyn and the comparison stars on the celestial plane are shown in Figure \ref{fig:comparison-star-position}.
C2 and C3 stars were used for observations in the late stage of the superoutburst when the C1 star was saturated because of a long exposure time.
The constancy of these stars was confirmed by observations over 13 nights between September 18, 2010 and November 6, 2010.
The differences in instrumental magnitude between these stars are $\Delta g'_\mathrm{inst}(\mathrm{C1-C2})=-1.548 \pm 0.001$ and $\Delta i'_\mathrm{inst}(\mathrm{C1-C2})=-1.373 \pm 0.001$ for the difference between C1 and C2 stars, and $\Delta g'_\mathrm{inst}(\mathrm{C1-C3})=-3.245 \pm 0.001$ and $\Delta i'_\mathrm{inst}(\mathrm{C1-C3})=-2.411 \pm 0.001$ between C1 and C3 stars. These values were used for the conversion of relative magnitudes between the object and each of the C2 and C3 stars, to that between the object and the C1 star.
%
Absolute photometry of the C1 star was performed on three photometric nights (November 1 and 24, 2010 and January 7, 2011).
We estimated that the standard system magnitudes of the C1 star are $g'=11.789 \pm 0.008$ and $i'=11.498 \pm 0.006$, derived from the weighted average of the results from the three nights.

The log of photometric observations is summarized in Table \ref{tab:photobslog}.
Note that our results are not corrected for galactic extinction, because extinction toward EZ Lyn is not significant ($E[B-V] = 0.049$ by \cite{schlafly+2011} via NED Galactic Extinction Calculator). Our conclusions and discussions in this paper were not influenced by correction of the extinction.
The barycentric Julian date of barycentric dynamical time, BJD${}_\mathrm{TDB}$, was calculated from UTC via http://astroutils.astronomy.ohio-state.edu/time/utc2bjd.html (see \cite{bjdtdb}).
The magnitude errors in Table \ref{tab:photobslog} do not include that of the C1 star because the error in C1 magnitude introduced a systematic rather than a random error to the standard system magnitudes of the object.
The orbital phase was calculated using the ephemeris of \citet{kato+2009b} with conversion from BJD${}_\mathrm{UTC}$ to BJD${}_\mathrm{TDB}$,

\begin{equation}
 \mathrm{Min}(\mathrm{BJD}_\mathrm{TDB}) = 2453799.3662(7) + 0.0590048(2)E.
\label{eqn:orbitalephemeris}
\end{equation}

\subsection{Spectroscopy}
The spectroscopy of EZ Lyn was performed on six nights between September 25, 2010 and October 16, 2010, using a low-resolution spectrograph, LOSA/F2 (Low-resolution Optical Spectrograph for the Araki telescope with F/2 optics, \cite{shinnaka+2013}), equipped on the Nasmyth focus of the Araki telescope.
This instrument had a slit (2.9 $\times$ 194 arcsec in the sky), grism (600 grooves/mm and the apex angle 37${}^\circ$), a Fe-Ne-Ar lamp for wavelength calibration, and a CCD camera (Apogee Alta U-47, e2v CCD47-10 1024$\times$1024 pixels with 13\,$\micron$/pixel,  gain $=$ 1.4\,e${}^{-}$/ADU, and read-out noise $\sim 10$\,e${}^{-}$).
The wavelength coverage was 380--780\,nm and the resolving power ($R \equiv \lambda/\Delta \lambda$) was about 580 at H$\alpha$ line.

The spectroscopic data were also reduced with IRAF, using the standard techniques for spectroscopy (e.g., dark-subtraction, flat-fielding, extraction of 1D spectra, wavelength calibration, summing the spectrum of each frame, sensitivity calibration).
The accuracy of the wavelength calibration (root mean square) was less than 0.02\,nm ($\sim 1/20$ pixel).

The log of spectroscopic observations is summarized in Table \ref{tab:specobslog}. 
The $\mathrm{BJD}_\mathrm{TDB}$ of the observations in this table was calculated using the same method as that for Table \ref{tab:photobslog}.

\section{Results}
\subsection{Light Curves of 2010 Superoutburst}
\begin{figure}
  \begin{center}
    \FigureFile(120mm,80mm){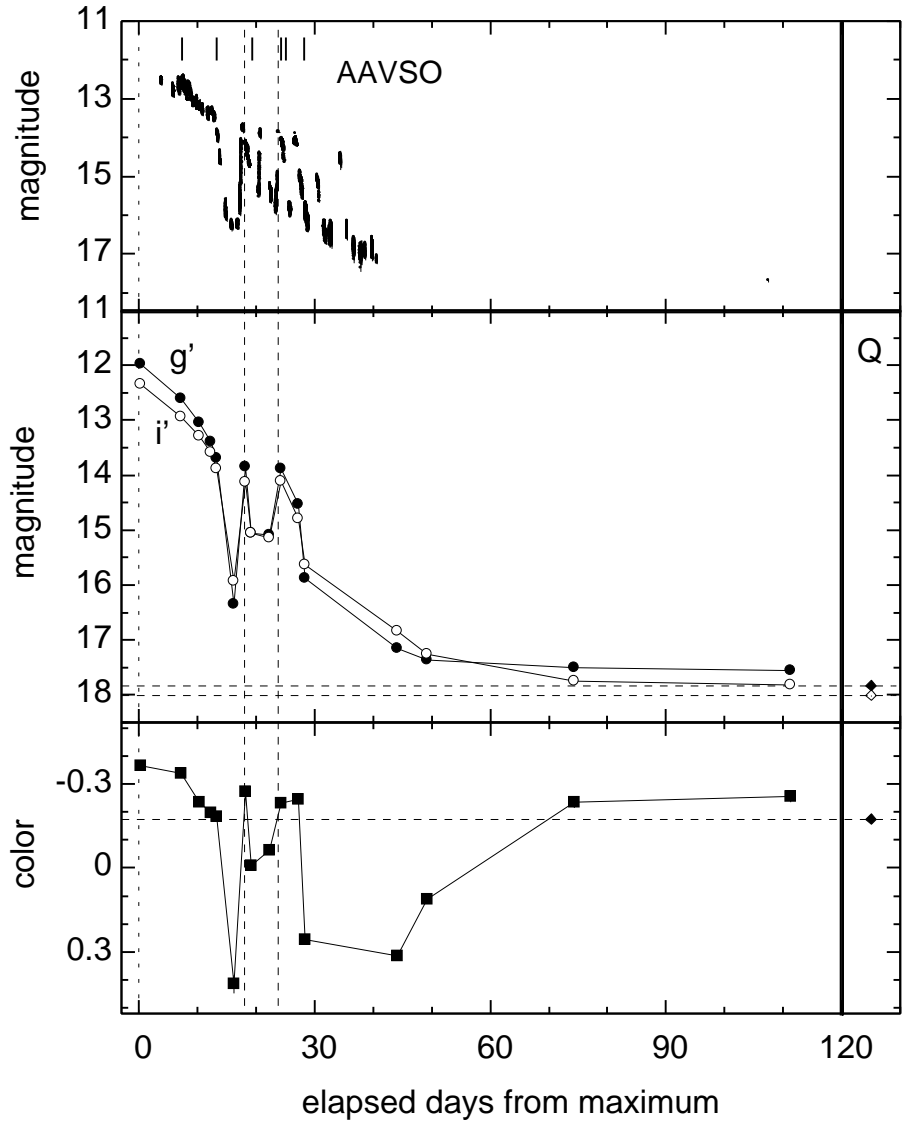}    
  \end{center}
  \caption{Light curves of EZ Lyn during the 2010 superoutburst. The top panel shows a light curve in the AAVSO International Database, and the middle and bottom panels are the light curves and the color variation ($g'-i'$) of our results. The filled and open circles in the middle panel represent $g'$- and $i'$- band magnitudes averaged for each night, respectively. `Q' indicates the magnitude and color in SDSS DR8, which we consider as a value from the quiescent phase. Vertical short lines in the top panel and long dashed lines indicate the timings of the spectroscopic observations and the observed rebrightening maxima, respectively.}\label{fig:lc_gi_ave1n}
\end{figure}

Figure \ref{fig:lc_gi_ave1n} shows light curves in the SDSS $g'$ and $i'$ bands, and $g'-i'$ color, from start to finish of the 2010 superoutburst, as well as a light curve from the American Association of Variable Star Observers (AAVSO) International Database.
This figure also plots the magnitude from the SDSS eighth data release (DR8) ($g$=$ 17.843 \pm 0.005$, $i$= $18.018 \pm 0.008$), which we consider the quiescent magnitude.
This outburst reached its maximum brightness on September 18, 2010 \citep{kato+2012}.
We defined the origin of the elapsed days from the maximum, $T$, as this day (BJD2455458.0).
The amplitude of the 2010 outburst was at least 5.9 mag for $g'$ and 5.7 mag for $i'$, which were the differences of magnitude between that on September 18, 2010 and SDSS DR8.
It should be noted that the brightness of this object did not return to quiescent levels even when $T=111$.

During the main superoutburst, the brightness decreased rapidly over time (from 0.089 and 0.085 mag $\mathrm{day}^{-1}$ at $T=0$--7 to 0.323 and 0.308 mag $\mathrm{day}^{-1}$ at $T=12$--13 for $g'$ and $i'$, respectively).
The color reddened with time and was at its reddest during the dip between the main superoutburst and rebrightening.
The color was bluer [$\Delta(g'-i')=-0.17$ at $T=22$--24] during the following rebrightening phase ($\Delta g'=-1.21$ at $T=22$--24).
In the slow declining phase ($T=42$--111) after the rebrightening, the color became bluer over time, more so than at the quiescent level.

\subsection{Temperature and size of emitting region}\label{section:temp+size_photometry}
In this section, we roughly estimate the temperature and size of the disk from the two-band photometric results.
In general, dwarf novae have four continuum sources: a white dwarf; an accretion disk; a secondary star; and a hot spot.
During an outburst, almost all of the optical emissions from a dwarf nova originate from the accretion disk (see \cite{warner1995}).
On the basis of the standard disk theory, the temperature varies with the radius, which is higher in the inner region and lower in the outer region. 
The continuum emission from the entire disk is a sum of blackbody radiations at rings with different radii, that is, with different temperatures.
Meanwhile, \citet{matsui+2009} recently reported that from multicolor ($g'VR_\mathrm{c}I_\mathrm{c}JK_\mathrm{s}$ bands) photometric results of the WZ Sge-type star V455 And, the continuum between the $g'$ and $J$ bands during the superoutburst can be reproduced by a single blackbody function.
Since the inner disk region with higher temperature ($T_\mathrm{in} \sim (3\mathrm{-}6) \times 10^4$ K) \footnote{These values were roughly estimated for WZ Sge-type objects, by assuming a standard steady-state temperature distribution in the accretion disk (Eq. 2.35 in \cite{warner1995}), the mass accretion rate as that of WZ Sge 2001 superoutburst ($10^{17}\mathrm{-}10^{18}$ g/s, \cite{patterson+2002}, \cite{long+2003}, \cite{godon+2006}), the mass of white dwarf $0.6 \MO$, the inner disk radius $r_\mathrm{in} = 49/36 R_\mathrm{WD} \sim 9 \times 10^9 $ cm, and the outer disk radius $0.6 A$ (2:1 resonance radius) $\sim 21 r_\mathrm{in}$, where $R_\mathrm{WD}$ and $A$ denote the radius of the white dwarf and the binary separation, respectively.}
emits most of its energy in the ultraviolet wavelength and the outer region with lower temperature ($T_\mathrm{out} \sim (0.6\mathrm{-}1) \times 10^4$ K)\footnotemark[1]{} emits in the optical wavelength, this result indicates that the optical continuum of WZ Sge-type objects during the superoutburst is emitted mostly from the outermost region of the disk having the lowest temperature in the disk.
Thus, we fitted $g'$ and $i'$ magnitudes with a single blackbody function and considered the derived temperature and size information as those of the outermost region of the disk, as an approximation.
The procedure for obtaining the temperature and size information was as follows. 
First, we calculated the energy of the radiation at $g'$ and $i'$ bands in flux from the observed magnitudes, and obtained the SED of EZ Lyn. Second, we fitted this SED with a single blackbody function, $f(\lambda) = a \lambda^{-5} \left ( \exp \left (b \lambda^{-1} \right) -1 \right )^{-1}$, where $\lambda$ is wavelength, $a$ is the size information that is proportional to the emitting area $A$ as $a \propto A d^{-2}$ ($d$ is the distance between the observer and the object), and $b$ is related to temperature $T_\mathrm{bb}$ by $b = hc(k_\mathrm{B}T_\mathrm{bb})^{-1}$, where $h$ is the Planck constant, $c$ is the speed of light, and $k_\mathrm{B}$ is the Boltzmann constant.
Finally, we calculated the temperature and size ratio of the emitting region from the best fit values of the free parameters $a$ and $b$.
Unfortunately, the errors in the free parameters could not be obtained by the above procedure because the fitting of only two data ($g'$, $i'$ magnitudes) with two free parameters had no degree of freedom ($\nu=0$).
To estimate the errors, we generated pseudo data (1,000) for $g'$ and $i'$ magnitudes using the Monte Carlo method, following Gaussian distributions whose averages and standard deviations were observed magnitudes and their errors, applied the above procedure to the pseudo data ($g_i'$ and $i_i'$ for $i=1,2,\cdot\cdot\cdot,1000$), calculated the average and standard deviation (strictly the square root of the unbiased variance) for the derived 1,000 parameter set ($a_i$ and $b_i$ for $i=1,2,\cdot\cdot\cdot,1000$), and considered the average and the standard deviation as the best fit value and the error of the parameters $a$ and $b$, respectively.
This result is consistent with those using 10,000 pseudo-data.

\begin{figure}
  \begin{center}
    \FigureFile(120mm,160mm){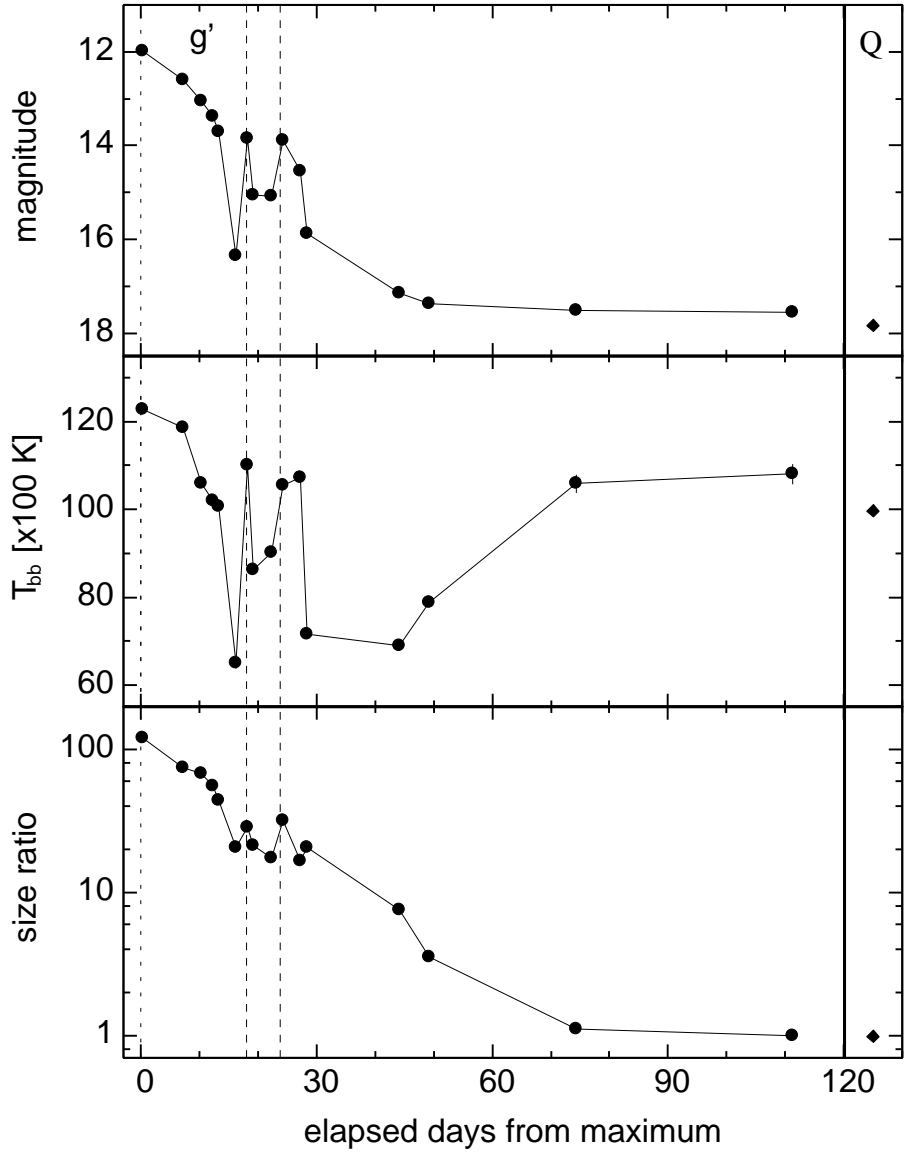}
  \end{center}
  \caption{Temporal evolution of the $g'$-band magnitude, the temperature and the size ratio of the emitting region. `Q' and vertical long dashed lines are the same as in Figure \ref{fig:lc_gi_ave1n}. The errors of the temperature and the size ratio are typically 40--$120K$ ($T=0$--49) and 0.06--2.0, respectively. For many data points, these are smaller than the mark size in the panels.}\label{fig:d-gTra}
\end{figure}

Figure \ref{fig:d-gTra} shows the temporal evolution of $g'$ magnitude, the temperature, and size ratio during the 2010 superoutburst. 
The size ratio is normalized to that at $T=111$ when the brightness is very close to the quiescent level.
During the main superoutburst ($T=0$--13) and rebrightening phases ($T=18$--28), the temperature and size of the emitting region decreased as brightness faded.
During the slow decline phase ($T=44$--111), the size decreased monotonically over time, whereas the temperature was at first very low but increased over time, and was finally hotter than at the quiescent phase.
At $T=111$, the size of the emitting region agreed with that during the quiescent phase.
We consider the evolution in the slow decline phase as an estimation, because our approximation of single blackbody radiation was not valid at this phase, and the additional contribution of emission sources, such as a hot spot was expected.

\subsection{Estimation of the disk radius ratio and comparison with a no-rebrightening system}\label{section:the_size_of_disk}
\begin{figure}
  \begin{center}
    \FigureFile(140mm,100mm){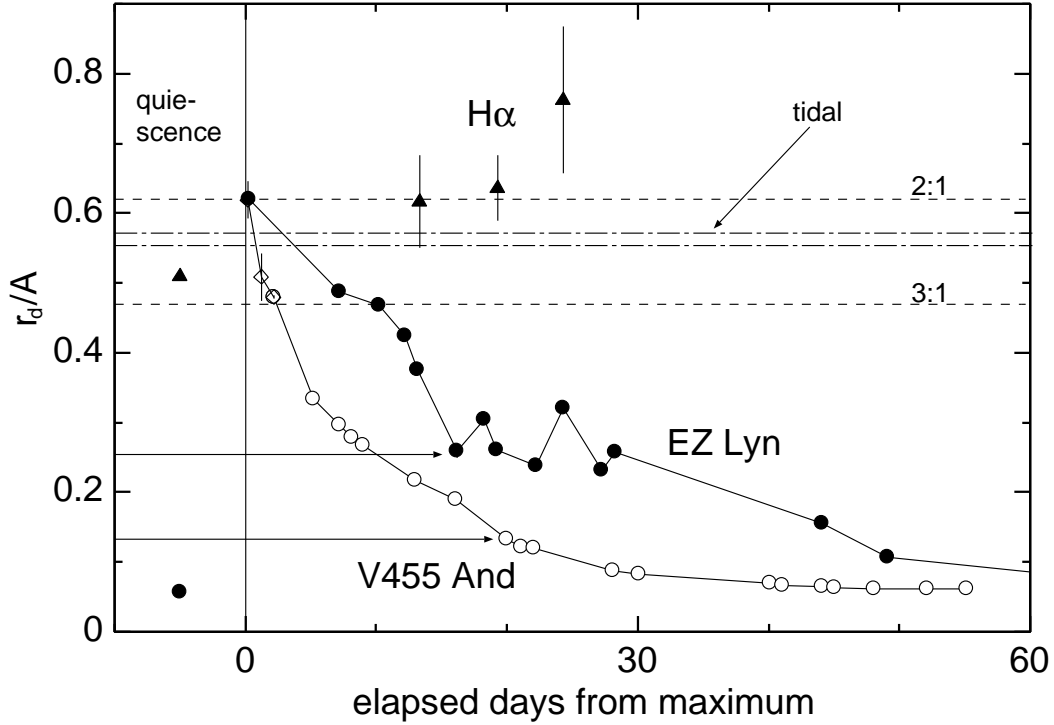}
  \end{center}
  \caption{Temporal evolution of disk radius ratio -- disk radius $r_\mathrm{d}$ per binary separation $A$ -- for WZ Sge-type objects EZ Lyn and V455 And. The radius ratio derived from the H$\alpha$ line profile is also plotted. The horizontal arrows point to the results when the main superoutburst ended. In addition, horizontal dot-dashed and dashed lines indicate tidal truncation (the upper one is for EZ Lyn and the lower for V455 And), 2:1 resonance, and 3:1 resonance radii, respectively. (See \cite{whitehurst+king1991}).}\label{fig:days_vs_disk_radius}
\end{figure}

In section \ref{section:temp+size_photometry}, we estimated the size ratio of the emitting region from $g'$- and $i'$- band magnitudes.
In this section, we calculate the ratio of the accretion disk radius $r_\mathrm{d}$ to the binary separation $A$ (disk radius ratio $r_\mathrm{d}/A$) of EZ Lyn from the estimated size ratio, and compare it to that of another WZ Sge-type object, V455 And, which was also fortunately observed from the rapid rising state \citep{matsui+2009} and did not show a rebrightening during the 2007 superoutburst.
For this calculation, we assumed that the size of the emitting region was proportional to the square of the disk radius, and that the disk radius ratio of both objects at the maximum was equal to that of 2:1 resonance.
The results are shown in Figure \ref{fig:days_vs_disk_radius}.
V455 And was observed with $V$, $J$, and $K_\mathrm{s}$ bands simultaneously and/or with $g'$, $R_\mathrm{c}$, and $I_\mathrm{c}$ bands simultaneously \citep{matsui+2009}.
To estimate the size of the emitting region of V455 And, we used only $g'$- and $I_\mathrm{c}$- band data in the same manner as EZ Lyn, because the $I_\mathrm{c}$ band had the most similar effective wavelength to the $i'$ band among the five bands.
At $T=0$ and 1, when V455 And was not observed in $g'$ and $I_\mathrm{c}$ bands, we used $V$- and $J$- band data. 
The size ratio during this period, $RA_{g'I_\mathrm{C}}(T)$ was scaled by the results at $T=2$ when photometry was carried out with both pairs, using the formula,
\begin{equation}
  RA_{g'I_\mathrm{C}}(T) = \frac{a_{VJ}(T)}{a_{VJ}(T=2)} \times RA_{g'I_\mathrm{C}}(T=2),
\end{equation}
where $a_{VJ}(T)$ is the size parameter of blackbody fit derived from a pair of $V$- and $J$- band data.
The derived disk radius is thought to represent that of the optically thick region in the accretion disk. 
In addition, the emissivity in the optically thick region depends only on the surface temperature at the emitting region.
Figure \ref{fig:days_vs_disk_radius} shows that the disk radius ratio of EZ Lyn during the main superoutburst ($T=0$--13) decreases more slowly ($0.019$\,day${}^{-1}$) than that of V455 And during the same phase ($0.027$\,day${}^{-1}$, $T=0$--16).
In addition, the disk radius ratio of EZ Lyn at the end of the main superoutburst (0.26 at $T=16$) is larger than that of V455 And (0.13 at $T=20$).
These results indicate that the accretion process in EZ Lyn works less effectively, and that there is a greater amount of gas left in the outermost region of the disk, than in V455 And.

There are two competing models to explain the rebrightening phenomenon: the mass reservoir model (\cite{kato+1998}; \cite{hellier2001}; \cite{osaki+meyer2001}); and the enhanced mass transfer (EMT) model (\cite{patterson+1998}; \cite{hameury+2000}).
As described in section 1, the EMT model predicts a {\it rapid} decrease in disk radius because the gas transferred from the secondary star has low angular momentum. 
In contrast, the mass reservoir model assumes that the rebrightening is caused by delayed accretion of gas left in the outermost region beyond the 3:1 resonance radius, and expects a {\it slow} decrease in disk radius and a large disk radius at the end of the main superoutburst.
Our results favor the mass reservoir model rather than the EMT model.

\subsection{Spectroscopy}
\begin{figure}
  \begin{center}
    \FigureFile(160mm,120mm){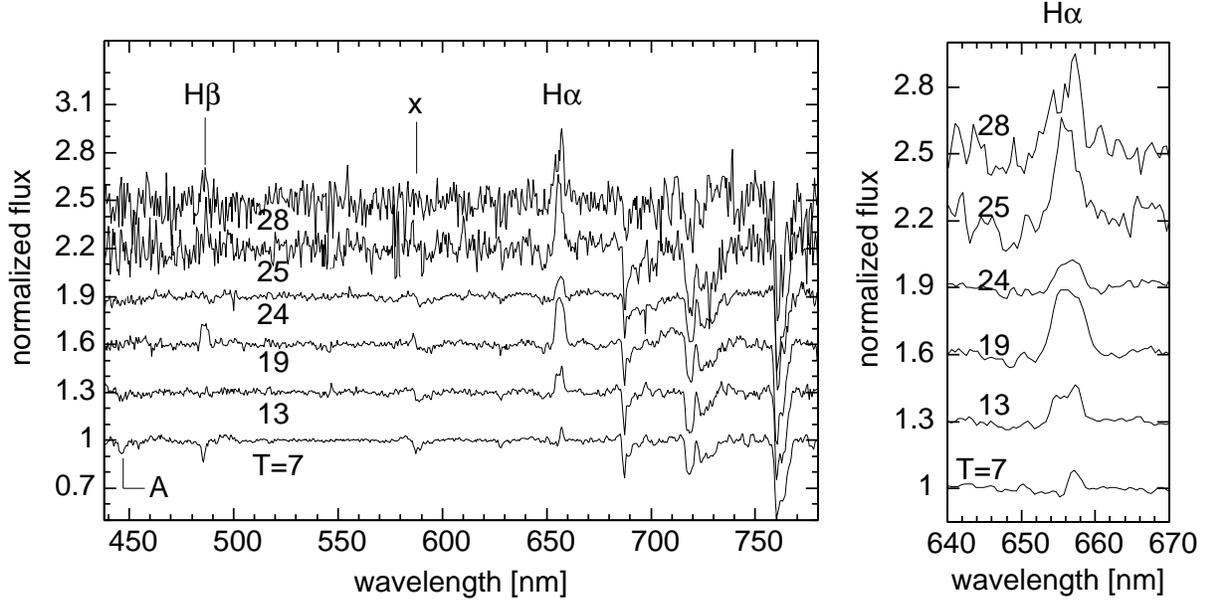}
  \end{center}
  \caption{Spectral evolution during the superoutburst. Left panel: low resolution spectra of the entire optical region (438--780\,nm). Right panel: enlarged view of the H$\alpha$ line region. Each spectrum is normalized to a unity continuum value, and is shifted by 0.3 to the vertical direction for visibility.  In the left panel, three lines are identified: H$\alpha$, H$\beta$, and the possible He\emissiontype{I} $\lambda$447.1\,nm (only when $T=7$, indicated by `A'). Note that the emission-like feature at $\lambda = 586$\,nm (`x') is not the He\emissiontype{I} line $\lambda$587.6\,nm but the residual of sky subtraction. Sharp emission- and absorption-like features at around 546\,nm and 578\,nm particularly seen at $T=25$ and $T=28$, are also the residual of sky subtraction, due to significant light pollution (mainly Hg).}\label{fig:spec_all+ha_zoom}
\end{figure}

Figure \ref{fig:spec_all+ha_zoom} shows the spectral evolution during the superoutburst.
Each spectrum is normalized to a unity continuum value, and is shifted by 0.3 to the vertical direction for visibility.
In the left panel H$\alpha$, H$\beta$, and possibly He\emissiontype{I} $\lambda$447.1\,nm lines are detected as an emission and/or absorption feature, whereas high ionization lines such as Bowen blend C\emissiontype{III}/N\emissiontype{III} around $\lambda$464\,nm, and He\emissiontype{II} $\lambda$468.6\,nm lines are not recognized.
In contrast, \citet{nogami2004} reports that He\emissiontype{II} and C\emissiontype{III}/N\emissiontype{III} lines are detected in WZ Sge 2001 superoutburst even during the mid-plateau phase, with peak intensity at about 0.1 to the continuum level.
This indicates that the EZ Lyn 2010 superoutburst was less energetic than the WZ Sge 2001 superoutburst.
According to \citet{kato+2009b}, EZ Lyn is a grazing, eclipsing system whose orbital inclination angle is similar to WZ Sge ($i=77\pm2^{\circ}$, \cite{steeghs+2007} and references therein).
According to \citet{warner1995}, the standard disk model indicates that the boundary layer (BL) has the highest temperature in the binary system, and the non-detection of high ionization lines indicates that the BL of EZ Lyn has a lower temperature than that of WZ Sge. 
The temperature of BL, $T_\mathrm{BL}$, depends on the mass of the white dwarf $M_\mathrm{WD}$, the radius of the boundary region $R$, and the mass accretion rate in the disk $\dot{M}_\mathrm{d}$ as $T_\mathrm{BL} \propto M_\mathrm{WD}^{1/3} R^{-7/9} \dot{M}_\mathrm{d}^{2/9}$ (Eq. 2.55b in \cite{warner1995}).
Considering these parameters, \citet{zharikov+2013} suggested that EZ Lyn has a massive white dwarf similar to the estimation of WZ Sge (\cite{steeghs+2007}). 
If these estimations are reliable, the low BL temperature of EZ Lyn is caused either by the disk having a cavity in the innermost region (large $R$), or by the mass transfer rate in the disk being smaller than that of WZ Sge (small $\dot{M}_\mathrm{d}$).

\begin{figure}
  \begin{center}
    \FigureFile(120mm,80mm){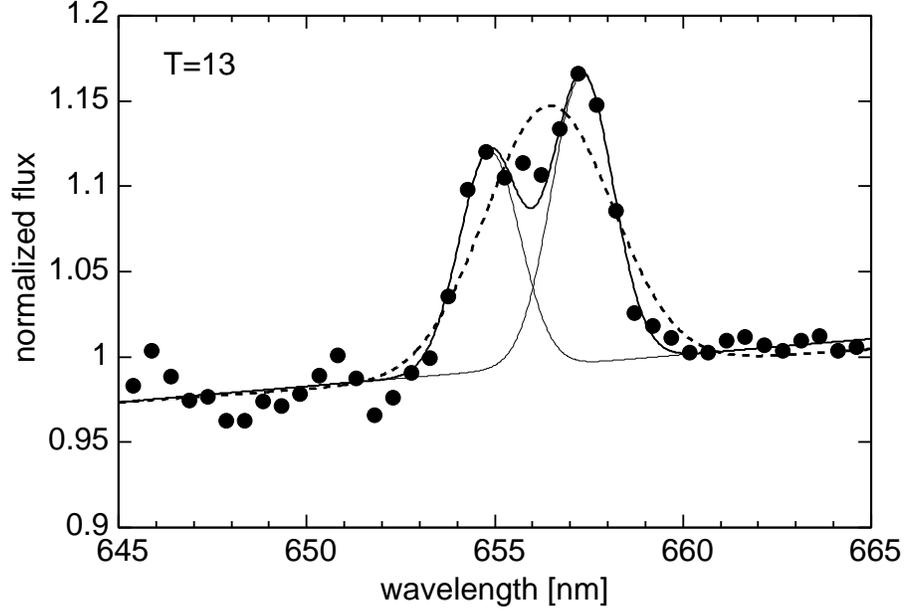}
  \end{center}
  \caption{Gaussian fitting of the H$\alpha$ line profile at $T=13$. The dashed curve shows the result of the single Gaussian fitting. The bold solid curve shows that of the two-Gaussian fitting, and the thin solid curves show each component of the two-Gaussian fitting. The chi-square $\chi^2$ and the degree of freedom $\nu$ between 653 and 660\,nm are $\chi^2/\nu = 55.1/11$ for the single Gaussian fitting and $7.87/8$ for the two-Gaussian one, respectively.}\label{fig:spec_Ha_profile_T13}
\end{figure}

The right panel in Figure \ref{fig:spec_all+ha_zoom} shows the line profile evolution of H$\alpha$.
The profile evolved from the combination of blue absorption and red emission components ($T=7$) to the broad and bell-shaped emission component ($T=13$--24).
In particular, the profile at $T=13$ clearly indicated a double, peak-like feature.
It is widely accepted that an emission line with a double-peak profile (e.g. \cite{horne+marsh1986}) is from an accretion disk with a large orbital inclination angle.
Thus, we tried to fit this profile with two Gaussian functions, and compared the results of the two functions with a single Gaussian curve fitted to the observed profile (Figure \ref{fig:spec_Ha_profile_T13}).
The derived chi-square $\chi^2$ and the degree of freedom $\nu$ between 653 and 660\,nm, are $\chi^2/\nu =55.1/11$ for the single Gaussian fitting and $\chi^2/\nu=7.87/8$ for the two Gaussian ones, respectively.
From Figure \ref{fig:spec_Ha_profile_T13} and $\chi^2/\nu$, we confirmed that the two-Gaussian curve can reproduce the observed profile much better than the single one.
Thus, it can be considered that emissions originating from the disk are dominant during the superoutburst, as in the quiescent phase (\cite{szkody+2006}).

To obtain the disk size information from H$\alpha$ line, we also fit the profiles observed on other days (in Table \ref{tab:specobslog}) and also in quiescence, obtained from the SDSS DR8 archive.
The results are summarized in Table \ref{tab:spec_Ha_gaussian_fits}.
We used the results to estimate the disk radius ratio derived from H$\alpha$ line in section \ref{section:disk_radius_ratio_Ha_line}.

\subsection{Early and ordinary superhumps}\label{section:early_ordinary_superhumps}
\begin{figure}
  \begin{center}
    \FigureFile(160mm,100mm){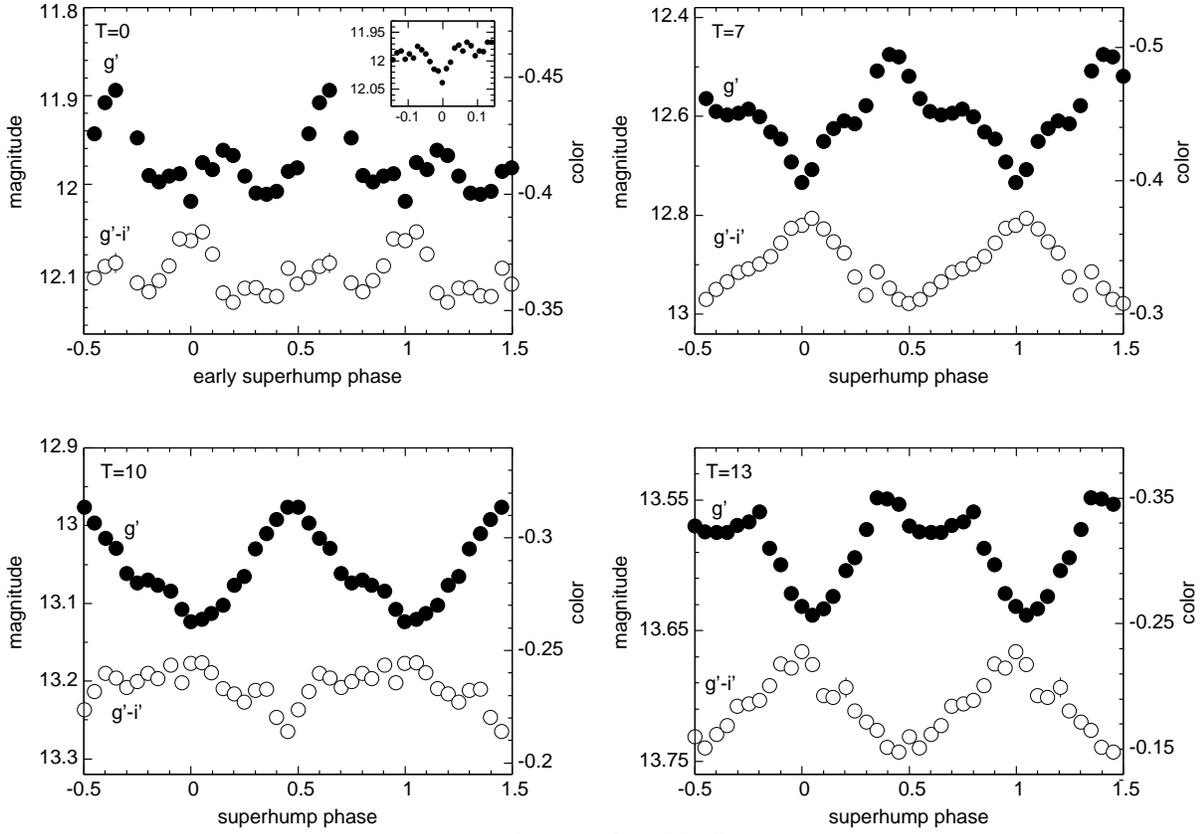} revise-version No.2
  \end{center}
  \caption{Phase-averaged light curves and color variations in early and ordinary superhumps. The top-left panel includes non phase-averaged $g'$-band light curve around the early superhump phase $=0$. The origin of early and ordinary superhump phases is at the minimum of $g'$-band brightness. The errors of the non-averaged $g'$ band, averaged $g'$ band, and $g'-i'$ color are typically 0.003, 0.001--0.003, and 0.002--0.004 mag, respectively. These are smaller than the mark size in the panels.}\label{fig:eshp+shp_vs_g+g-i_T0-13}
\end{figure}

At $T=0$, 7, 10, and 13, EZ Lyn was observed by high-quality continuous photometry during approximately one orbital period ($T=0$) and over two orbital periods ($T=7$, 10, and 13). 
This photometry enabled us to study the color variations caused by early and ordinary superhumps.
To see small variations, we calculated phase-averaged light curves for each night.
Figure \ref{fig:eshp+shp_vs_g+g-i_T0-13} shows the phase-averaged light curves and the color variations of both early and ordinary superhumps.
In this figure, filled and open circles display $g'$ magnitude and $g'-i'$ color, respectively.
The observed light curve at $T=0$ is folded with the early superhump period of 0.058972d, reported in Kato et al. (2010a)\footnote{vsnet-alert 12196 (http://ooruri.kusastro.kyoto-u.ac.jp/mailarchive/vsnet-alert/12196)}, whereas those at $T=7$, 10, and 13 are folded with the superhump periods of 0.05932d (Kato et al. 2010b)\footnote{vsnet-alert 12201 (http://ooruri.kusastro.kyoto-u.ac.jp/mailarchive/vsnet-alert/12201)}, 0.059496d (Kato et al. 2010c)\footnote{vsnet-alert 12210 (http://ooruri.kusastro.kyoto-u.ac.jp/mailarchive/vsnet-alert/12210)}, and 0.05981d (Kato et al. 2010d)\footnote{vsnet-alert 12218 (http://ooruri.kusastro.kyoto-u.ac.jp/mailarchive/vsnet-alert/12218)}, respectively.
Since the observed light curve at $T=13$ shows a gradual decrease trend superimposed on short-term variations, this decreased component was detrended by subtracting a linear fitted function before folding.
The top left panel in Figure \ref{fig:eshp+shp_vs_g+g-i_T0-13} shows the light curve and color variation at $T=0$.
This panel clearly displays two humps during one period in the $g'$ light curve, which is typically seen in the early superhump phase.
The $g'$ brightness decreased at phase $=0$, which can be recognized as a sharp dip in the non phase-averaged $g'$-band light curve (upper right graph inside the panel).
The timing of this dip corresponds with the minimum of the ephemeris of \cite{kato+2009b} (Eq.\,(\ref{eqn:orbitalephemeris}) in this paper).
Around this time, the object was at its bluest.
However, the duration of the $g'-i'$ color becoming bluer is longer than the dip width in the $g'$-band light curve.
Thus, this color variation was {\it not} mainly caused by the occultation of the outermost region of the disk by the secondary star, although this object was considered as a grazing eclipse system (\cite{kato+2009b}).
Except for the dip feature, this figure also reveals that the color becomes blue around the early superhump minimum ($\Delta(g'-i')=-0.025$ around phase$\sim 0$) and is redder at the two humps (phase $\sim 0.2$ and $\sim 0.65$).
These color behaviors are roughly the same as those seen in V455 And and in OT J012059.6+325545 \citep{nakagawa+2013}.
The amplitude of the color variation of EZ Lyn (with the amplitude of the brightness $\Delta g' \sim 0.11$ mag) is smaller than that of V455 And ($\Delta(g'-I_\mathrm{c}) \sim 0.06$ with $\Delta g' \sim 0.15$ mag at $T=5$), but is similar to that of OT J012059.6+325545 ($\Delta(g'-i') \sim 0.02$ with $\Delta g' \sim 0.08$ mag at $T=1$, \cite{nakagawa+2013}).
Note that a periodicity of the color variation was not confirmed by short duration of our observation covering approximately one orbital period.
On the basis of both color behaviors and amplitudes, however, we can conclude that EZ Lyn is the third system where the color variation was successfully observed in early superhumps, next to V455 And \citep{matsui+2009} and OT J012059.6+325545 \citep{nakagawa+2013}.

The top right and bottom panels in Figure \ref{fig:eshp+shp_vs_g+g-i_T0-13} show the light curves and color variations for the ordinary superhump.
The object was bluest at the superhump minimum.
As the object increased in brightness, the color became redder. 
Although the profile of the $g'$-band light curve changes over time, these color behaviors are common in $T=7$--13.

On the basis of the folded color variation, we studied the physical property of the superhump source, using a similar method as that adopted by \citet{matsui+2009}.
If the early and ordinary superhumps are located at the outer part of the accretion disk, the system should have at least two emission components (inner constant and outer variable).
However, it was difficult for our data to separate both components without any assumptions.
Thus, we performed an analysis with the simplest model, the single blackbody emission component model, the same as in section \ref{section:temp+size_photometry}.
This model assumes that the superhump is caused by the entire disk.
We also performed the analysis with a two-component model under several assumptions. 
The result is shown in section \ref{section:two_components_fit}.

\begin{figure}
  \begin{center}
    \FigureFile(160mm,180mm){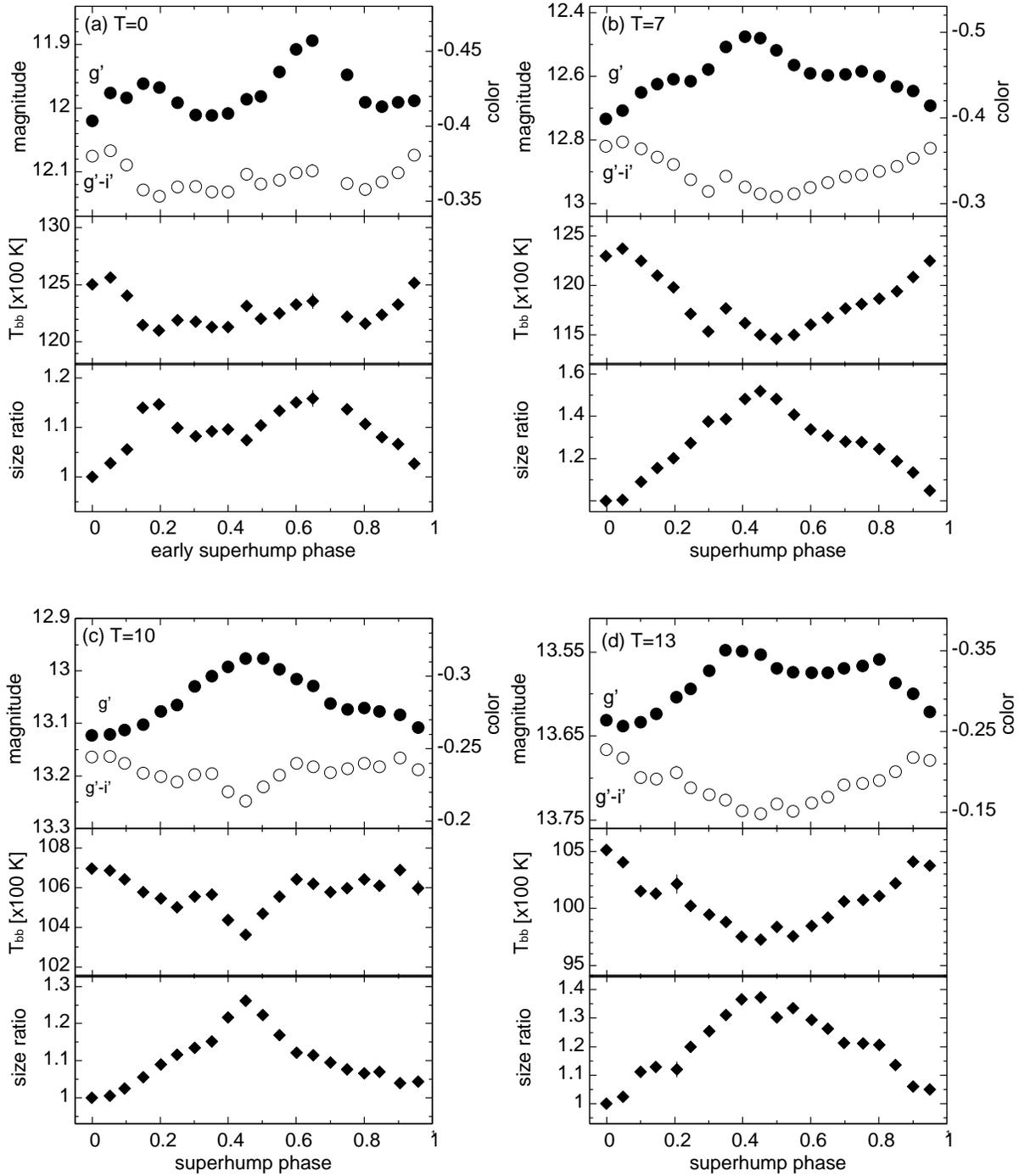}
  \end{center}
  \caption{Temporal variations in the temperature and size of the emission region derived from a single blackbody fitting, for the early and ordinary superhump phase. The top panels show the $g'$-band light curves and $g'-i'$ color variations.  The middle and bottom panels show the temperature and size of the blackbody emission region, respectively. The size was normalized by that at minimum $g'$ brightness. The abscissa indicates the early and ordinary superhump phase.}\label{fig:eshp+shp_vs_g+ra+T_T0-13}
\end{figure}

Figure \ref{fig:eshp+shp_vs_g+ra+T_T0-13} shows the temporal variations in the parameters of our model for early and ordinary superhumps.
The top panels display the folded $g'$-band light curves and $g'-i'$ color variations.
The middle and bottom panels display the estimated temperature and the size of the blackbody emission region, respectively.

Figure \ref{fig:eshp+shp_vs_g+ra+T_T0-13}a shows the result for the early superhump.
From approximately the minimum to the first hump (phase between $0.80$ and 0.10), the temperature was higher and the emitting size smaller than those at the hump maximum.
In addition, the $g'$-band light curve correlated well with the emitting size.
These properties are roughly the same as those seen in V455 And (\cite{matsui+2009}), which supports the idea that {\it the early superhump light source was a vertically expanded low-temperature region at the outermost part of the disk} (\cite{matsui+2009}).

Figure \ref{fig:eshp+shp_vs_g+ra+T_T0-13}b--d displays the results for the ordinary superhump.
The temperature decreased as the emitting size increased, was lowest around the phase when the emitting size was at its maximum, and increased as the size decreased.
This clear anticorrelation indicates that the superhump source has a lower temperature than other regions in the disk, and that the superhump is caused by the expansion of this low temperature region.

\section{Discussion}

\subsection{Effect of assumptions on our calculation of the disk radius ratio}
In section \ref{section:the_size_of_disk}, we calculated the disk radius ratios of both EZ Lyn and V455 And from the size parameters of blackbody fit.
For this calculation, we assumed that the size of the emitting region $S(r_\mathrm{d})$ is proportional to the square of the disk radius ($S(r_\mathrm{d}) \propto r_\mathrm{d}^2$), and that the disk radius ratio at the maximum is equal to that of the 2:1 resonance.
In this section, we discuss the effect of these assumptions on the results of the comparison between EZ Lyn and V455 And.
First, the assumption that the size depends on the disk radius does not affect the results, if both objects have the same function for the size $S(r_\mathrm{d})$.
This is because the order in size between both objects is unchanged under the same function.
Second, we assumed that the disk radius ratio at its maximum, $r_\mathrm{d,max}/A$, was the same between both objects and is equal to that of the 2:1 resonance radius, $r_{2:1}/A$.
Even if the disk radius ratio at its maximum differed from that of the 2:1 resonance radius, our result would not be affected by this assumption, as the disk radius ratio of both objects, at maximum, is the same.
It is possible that the maximum disk radius ratio differs between the two objects.
However, since the disk radius ratio of V455 And at the end of the main superoutburst is about half that of EZ Lyn, to create the condition that V455 And has a larger disk radius ratio than EZ Lyn, requires the disk radius of V455 And (at the maximum brightness) to exceed the binary separation, $r_\mathrm{d,max} \sim 2r_{2:1} > A$, which is very unlikely to occur.
Consequently, we concluded that our results in section \ref{section:the_size_of_disk} were not affected by the assumptions.

\subsection{Effect using $V$ and $J$ bands on the variation of the disk radius ratio of V455 And}\label{section:disk_radius_ratio_jk_v455and}
In section \ref{section:the_size_of_disk}, we used $V$ and $J$ bands instead of $g'$ and $I_\mathrm{c}$ to estimate the disk radius ratio of V455 And at $T=0$--1.
In this section, we discuss the effect of this on the variation of the disk radius ratio of V455 And and on the comparison of this ratio with that of EZ Lyn.
If the disk is an ideal blackbody emitter with a uniform temperature, the result derived by $V$ and $J$ bands would be the same as that by $g'$ and $I_\mathrm{c}$ bands.
In practice, however, since the disk has a multiple temperature structure, it is expected that the results by $V$ and $J$ bands would differ from that by $g$ and $I_\mathrm{c}$ bands.
The comparison of results by $V$ and $J$ with those by $g'$ and $I_\mathrm{c}$ bands at $T=2$--5, when all of these bands are observed, reveals that the amplitude of the variation of $g'$ and $I_\mathrm{c}$ bands is 15\% larger in size ratio and 7\% larger in disk radius ratio than that of $V$ and $J$ bands.
If this result holds for the variation at $T=0$--2 with the same rate for $T$, the disk radius of $g'$ and $I\mathrm{c}$ bands at $T=0$ is about 5\% larger than that estimated by Eq.\,(2).
This consequently means that the speed of decrease of the disk radius ratio of V455 And is about 5\% more rapid than that portrayed in Figure \ref{fig:days_vs_disk_radius}, and that the disk radius ratio at the end of the main superoutburst is also about 5\% smaller than that in Figure \ref{fig:days_vs_disk_radius}. This increases the differences in both values between EZ Lyn and V455 And.
Therefore, we concluded that the usage of $V$ and $J$ bands does {\it not} affect our conclusion in section \ref{section:the_size_of_disk}.

\subsection{The disk radius estimated by H$\alpha$ line profile}\label{section:disk_radius_ratio_Ha_line}
The separation of the two peaks in the H$\alpha$ line profile also enables us to estimate the disk radius \citep{warner1995}. 
This estimation requires the white dwarf mass $M_\mathrm{WD}$, orbital period $P_\mathrm{orb}$, orbital inclination angle $i$, and the mass ratio of the secondary star to the white dwarf $q$.
Since the eclipse feature was shown in the light curve during the superoutburst but was not clearly shown during the quiescent phase, we adopted $i \sim 75^\circ$, which is a similar angle to WZ Sge ($i=77 \pm 2^{\circ}$, \cite{steeghs+2007} and reference therein).
For the mass ratio $q$, we adopted $q = 0.056$, estimated by the relation between the superhump excess $\epsilon$ and the mass ratio \citep{patterson+2005} and by $\epsilon = 0.011$ (\cite{kato+2012}). 
The orbital period of EZ Lyn is $P_\mathrm{orb} = 0.0590048$\,d derived by \citet{kato+2009b}.
Furthermore, we assume $M_\mathrm{WD} = 0.6 \MO$, a typical mass for a white dwarf.
The disk radius ratio $r_\mathrm{d}/A$ is
\begin{eqnarray}
\frac{r_\mathrm{d}}{A} &=& 0.62 \left ( \frac{M_\mathrm{WD}}{0.6\MO} \right )^{2/3} \left ( \frac{\sin{i}}{\sin{75^\circ}} \right )^2 \left ( \frac{v_\mathrm{obs}}{560\mathrm{km/s}} \right )^{-2} \nonumber \\
&& \hspace{9em} \times \left ( \frac{1+q}{1.056} \right )^{-1/3}  \left ( \frac{P_\mathrm{orb}}{0.0590048\mathrm{d}} \right )^{-2/3},
\end{eqnarray}
which is derived from the combination of equilibrium between gravitation and centrifugal force,  
\[
G \frac{m M_\mathrm{WD}}{r_\mathrm{d}^2} = \frac{m}{r_\mathrm{d}} \left ( \frac{v_\mathrm{obs}}{\sin{i}} \right )^2, 
\]
and Kepler's third law, 
\[
A^3 = \frac{G}{4\pi^2} M_\mathrm{WD} (1 + q) P_\mathrm{orb}^2 ,
\]
where $G$ is the gravitational constant.
Eq.\,(3) indicates that the disk radius ratio increases with $M_\mathrm{WD}$ and/or with $i$.
The estimated disk radius ratio was also plotted in Figure \ref{fig:days_vs_disk_radius}.
Under this assumption, the disk radius ratio by H$\alpha$ line at $T=13$ was estimated as 0.62, larger than that of the photometric disk on the same day (0.38).

It is conceivable that the accretion disk of optically thin gas was more widely extended within the Roche-lobe than that of optically thick gas.
The disk radius ratio by H$\alpha$, however, depends strongly on $M_\mathrm{WD}$ and $i$.
For example, it requires $M_\mathrm{WD}=0.3\MO$ or $i=50^\circ$ from Eq.\,(3) to make the size of the H$\alpha$ disk at $T=13$ equal to the photometric one.
Such a low inclination is difficult to accept because EZ Lyn shows an eclipse feature in the light curve during the superoutburst \citep{kato+2009b}, which strongly suggests $i > 60^\circ$.
In addition, such a small mass for the white dwarf in EZ Lyn is also unrealistic because \citet{zharikov+2013} suggests that EZ Lyn has a massive white dwarf ($> 0.7\MO$) on the basis of the SED model fitting to the optical and near-infrared observation.
If the H$\alpha$-disk radius truly has a larger size than the photometric one, it would be a consequence that optically thin gases in this H$\alpha$ disk are associated with the matter remaining in the outermost region of the accretion disk and are responsible for the rebrightening phenomenon.
Furthermore, if that is true, the result that the H$\alpha$ disk remained constant in size during $T=13$--24 can be explained by gas in the outermost region remaining in abundance and being responsible for rebrightening continuing to at least $T=36$ (Figure \ref{fig:lc_gi_ave1n}).

\subsection{Difference of the temporal variation of the disk radius ratio between EZ Lyn and V455 And}
With regard to the photometric disk radius, V455 And's (no rebrightening system) disk more rapidly decreases in size over time than that of EZ Lyn with rebrightening.
This indicates that accretion from the outer to the central region occurred more effectively in V455 And than in EZ Lyn.
 There are two possible reasons: (1) the viscosity in the hot state is higher in V455 And; or (2) the angular momentum of gas in the outermost disk region is more effectively extracted by the secondary star in V455 And.
Since the source of viscosity is not clearly understood, we do not consider (1).
The physical properties of both systems are summarized in Table \ref{tab:system_info_V455And_vs_EZLyn}.
As shown in this table, V455 And has a larger $q$ and smaller $P_\mathrm{orb}$ than EZ Lyn.
Since it is considered that the tidal extraction of angular momentum from the disk by the secondary star is more effective in the system with larger $q$ and smaller $P_\mathrm{orb}$ \citep{hellier2001}, the second possibility seems more plausible, although the first is not excluded.
We note that we have only two samples; more are required to discuss the cause of effective extraction of angular momentum.

\subsection{Superhump source: contribution of constant component}\label{section:two_components_fit}
\begin{figure}
  \begin{center}
    \FigureFile(140mm,140mm){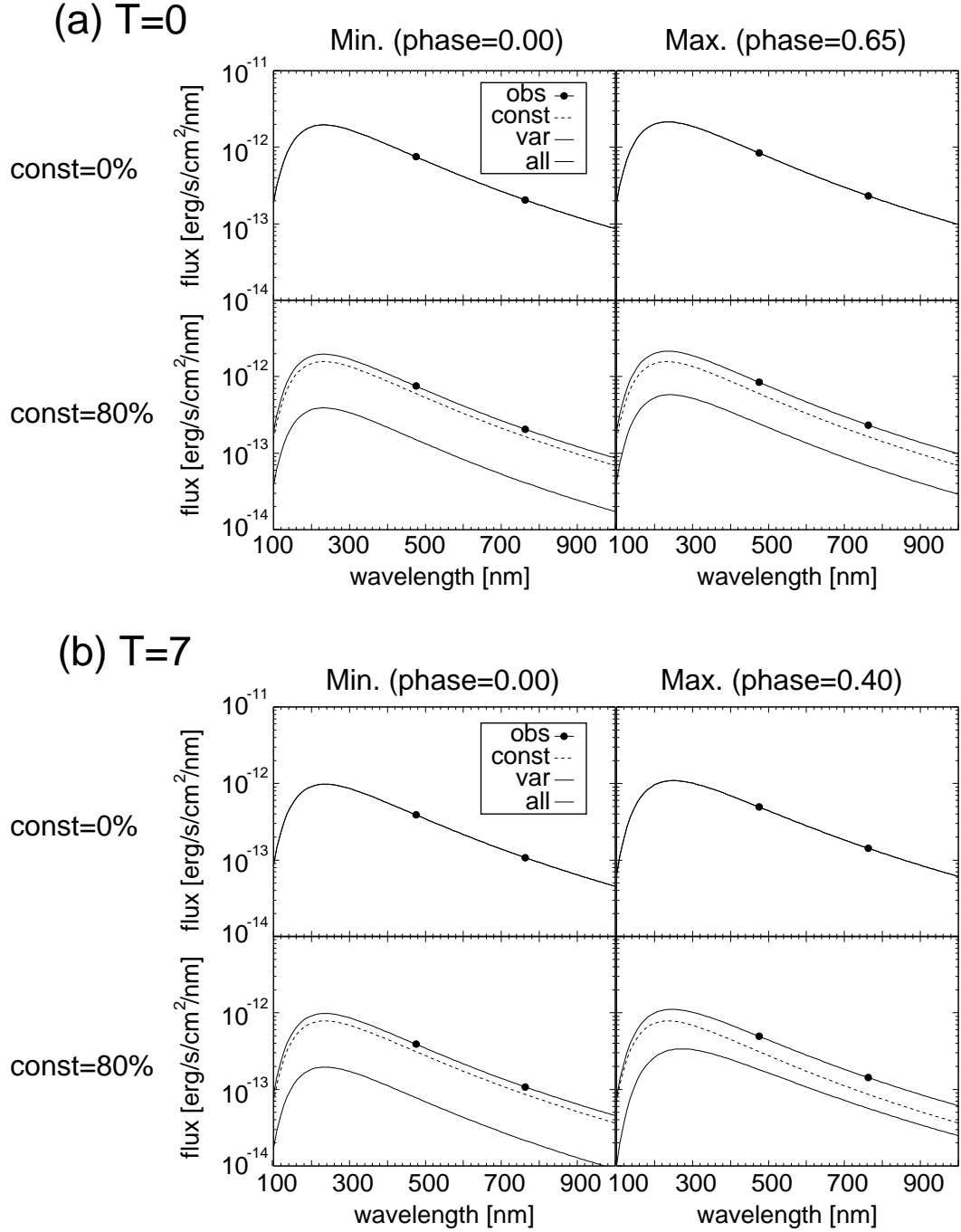}
  \end{center}
  \caption{Examples of the observed SEDs and the SEDs of the two component models. The filled circles indicate the observed fluxes in the $g'$ and $i'$ bands. The dashed thin solid, and thick solid lines show models with constant, variable, and both components, respectively.  Panel (a) shows the SEDs at $T=0$ when the early superhumps were observed, and Panel (b) those at $T=7$ when the ordinary superhumps were observed. In both panels, the left side presents the SEDs at the superhump minimum (superhump phase $= 0.00$), and the right side those at the superhump maximum (the early superhump phase $= 0.65$, and the ordinary superhump phase $= 0.40$). In both (a) and (b), the upper panels show the models with a constant component of $0\%$ of the total flux at the superhump minimum, and the lower ones show those with a constant component of $80\%$ of the total flux at the minimum.}\label{fig:wa_vs_flux_const0+80p_T0+7_2comp}
\end{figure}

In section \ref{section:early_ordinary_superhumps}, we modeled one emission component for early and ordinary superhumps, which means that humps represent changes in the temperature and size of the entire disk.
In this section, we consider a model of two emission components, variable and constant ones, in the same manner as \citet{matsui+2009}, and discuss the effect of the constant component on the results.
We assume that both constant and variable components have the same temperature at the hump minimum, and the temperature of the constant component is not changed during superhumps.
Under this assumption, we fit the model for early and ordinary superhumps with the flux of the constant component contributing 40\% and 80\% of total flux at the hump minimum.
The formula for the SED of this two-component model is
$f_\mathrm{all} (\lambda,\phi) = R \times B_\mathrm{const}(T_\mathrm{0},\lambda,\phi=0) + B_\mathrm{var}(T_\mathrm{bb},\lambda,\phi)$, 
where $\phi$ denotes the superhump phase, $R$ is the ratio of the flux of the constant component to the total flux at the hump minimum, and $T_\mathrm{0}$ is the temperature at the hump minimum for the one component model.
Figure \ref{fig:wa_vs_flux_const0+80p_T0+7_2comp} displays the examples of the observed and model SEDs. The panels in (a) are the SEDs at $T=0$ when the early superhumps were observed, and the panels in (b) are those at $T=7$ when the ordinary superhumps were observed. In these panels, the left ones display the SEDs at the hump minimum, the right ones those at the hump maximum, the upper ones those of the models with the constant component contributing 0\%, and the lower ones are those with it contributing 80\%.
We note that the 0\% model SEDs are the same as those of the one component model reported in section \ref{section:early_ordinary_superhumps}.

\begin{figure}
  \begin{center}
    \FigureFile(160mm,180mm){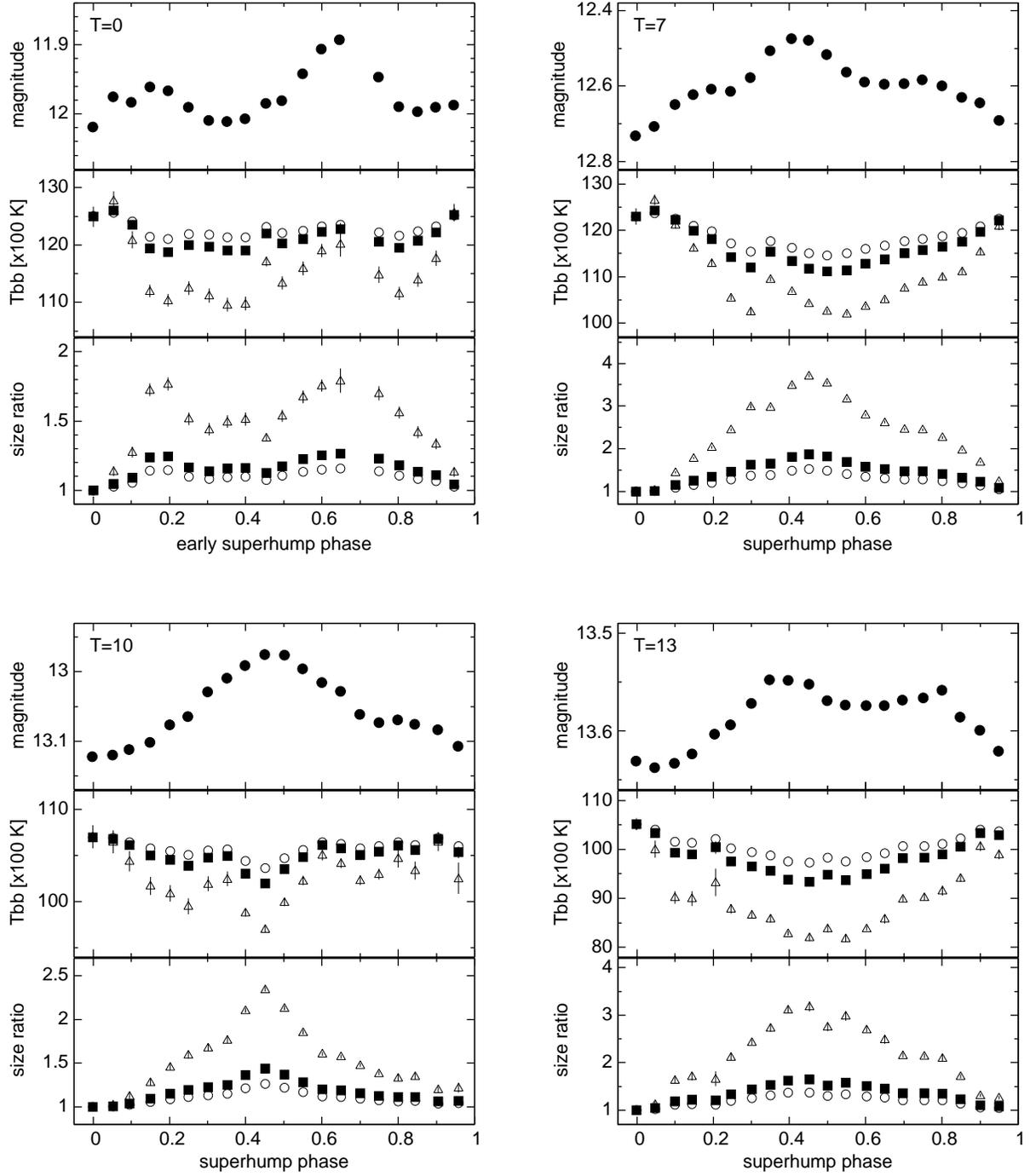}
  \end{center}
  \caption{Temporal variations of the temperature and the size of the emission region derived from two component (constant and variable) blackbody fit. The results for both early and ordinary superhumps are displayed. The top panels show the $g'$-band light curves.  The middle and bottom panels show the temperature and size of the blackbody emission region. The open circles, filled squares and open triangles are those with the contribution of the constant component 0\%, 40\%, and 80\% at the phase of minimum, respectively. The size was normalized by that at the minimum of $g'$ brightness.}\label{fig:eshp+shp_vs_g+ra+T_T0-13_2comp}
\end{figure}

Figure \ref{fig:eshp+shp_vs_g+ra+T_T0-13_2comp} shows the temporal variation of the model parameters for the hump component. 
The top, middle and bottom panels show the same as Figure \ref{fig:eshp+shp_vs_g+ra+T_T0-13}, except for excluding the color variation in the top panel.
The open circles, filled squares and open triangles plot the results of constant components contributing 0\%, 40\%, and 80\% of the total flux at the $g'$ hump minimum, respectively. 
Although models with a larger contribution of the constant component showed a larger amplitude in variations in both temperature and size, its profile of variations was qualitatively the same in all cases. 
This was also concluded by \citet{matsui+2009}.

From the amplitude of the size, we roughly estimated the thickness of the disk at the superhump source, using a simple model of two dimensions. 
In the case of the 80\% contribution model, the size expanded 1.8-fold, and over 3-fold from the hump minimum for the early and ordinary superhumps, respectively.
Early superhumps are proposed to be variations caused by vertical deformation of the disk \citep{kato2002}.
In this case, we can estimate the height, $h$, at the outermost region of the disk required to reproduce the 1.8-fold expansion of the apparent emitting area.
For this estimation, we suppose the accretion disk to take cylindrical form, and consider that we see a flat part ($h_\mathrm{min}=0.1r$, where $r$ is the disk radius) of the disk with an inclination of $75^\circ$ at the hump minimum.
The emitting size at the hump minimum is described by $S_\mathrm{min}= \pi r^2\cos{i} + 2rh_\mathrm{min}\sin{i}$, where the first term is the projected area of the disk surface and the second term is the projected cross-sectional area of the lateral face of the disk.
At the hump maximum, we see a vertically expanded part whose emitting size is $S_\mathrm{max}=\pi r^2\cos{i} + 2rh_\mathrm{max}\sin{i}$.
With this simple model and with the result $S_\mathrm{max}/S_\mathrm{min}=1.8$, we calculated the height to be $h_\mathrm{max} \sim 0.52r$ at the hump maximum.
This height is too large to be explained within the framework of the standard accretion disk model ($h \sim 0.1r$: \cite{sakura+sunyaev1973}). 
However, the 0\% contribution model yields $h_\mathrm{max} \sim 0.18r$, slightly higher than the standard model.
These results support the indication discussed in \citet{matsui+2009}; the contribution of the constant component in the inner disk should be small in the optical regime.

\subsection{Differences in the color variation of ordinary superhumps between EZ Lyn and V455 And}
As shown in section \ref{section:early_ordinary_superhumps}, the color variation in the ordinary superhumps of EZ Lyn is clearly anticorrelated with brightness.
This color variation is different from that of V455 And; EZ Lyn is reddest at minimum brightness, becomes bluer with increasing brightness, reaches its bluest before its peak of brightness, and reddens with a further increase in brightness.
\citet{matsui+2009} interpreted the variation in V455 And; at the brightening phase of the superhump, the increase in disk temperature is caused by a viscous heating effect. After the heating stopped, the object entered an expansion-cooling phase.

As a possible cause of the color-variation difference, there are several system parameters, including the orbital inclination angle $i$ and disk radius ratio $r_\mathrm{d}/A$.
Among these parameters, $i$ can be rejected because \citet{araujo+2005} estimated the orbital inclination of V455 And as $i \sim 75^\circ$, on the basis of the presence of shallow eclipses in its quiescent light curve, which is almost the same as that of EZ Lyn.
The mass ratio and orbital period are also not critically different between the systems, as summarized in table \ref{tab:system_info_V455And_vs_EZLyn}.
On the other hand, the photometric disk radius (when ordinary superhumps are observed) is clearly different; EZ Lyn has a larger disk radius ratio ($r_\mathrm{d}/A \sim 0.5$) than V455 And ($\sim 0.3$).
A larger disk radius ratio is thought to make the timescale of heating and expansion longer, which may cause the difference in color variation.

The color variation in the superhumps is a good tool to study the superhump light source.
However, as far as we know, there are only four examples of such a color study performed by high accuracy multi-color photometric observation using CCD camera on SU UMa-type dwarf nova: WX Cet (\cite{howell+2002}); V455 And (\cite{matsui+2009}); SU UMa (superhump-like modulation seen in the normal outburst, \cite{imada+2012} and \cite{imada+2013}); and EZ Lyn.
The number of samples should be increased through further observation.

\section{Summary}

We performed optical dual-band photometry and spectroscopy for the WZ Sge-type dwarf nova EZ Lyn during the 2010 superoutburst. 
Our studies are summarized as follows:

\begin{itemize}
\item Dual-band photometry revealed that the color $g'-i'$ was redder with the object being fainter during the main superoutburst and following the rebrightening phase ($T=0$--28). During the slow declining phase ($T=42$--111) after the rebrightening, the color became bluer over time and bluer than in quiescence.

\item We estimated the temperature and size ratio of the emitting region from $g'$- and $i'$- band data using a single blackbody function, and calculated the disk radius ratio from the size ratio, with noted assumptions.
Comparison of the temporal evolution of the disk radius ratios between EZ Lyn and V455 And (no rebrightening system) revealed: (1) the decrease speed of the disk radius ratio of EZ Lyn was slower than that of V455 And; and (2) the radius ratio of EZ Lyn just at the end of the main superoutburst was larger than that of V455 And.
These results indicate that the accretion process in EZ Lyn was less effective than in V455 And, and a greater amount of gas was left in the outermost region of the disk than in V455 And, which favors the mass reservoir model for the mechanism of rebrightening.

\item From spectroscopy during both the superoutburst plateau and the subsequent rebrightening phase, H$\alpha$, H$\beta$, and possibly He\emissiontype{I} $\lambda$447.1\,nm lines were detected, whereas high ionization lines such as He\emissiontype{II} $\lambda$468.6\,nm and Bowen blend C\emissiontype{III}/N\emissiontype{III} around $\lambda$464\,nm were not detected.

\item The H$\alpha$ emission line shows a double-peak profile as seen in the quiescent phase.

\item We estimated the disk radius ratio from the H$\alpha$ line profile with several assumptions, and compared it to that derived by photometry.
The comparison reveals that the H$\alpha$ disk radius ratio was larger than that derived by photometry.
This result suggests that optically thin gas was extended to the outer region further than the optically thick (photometric) disk.
This optically thin gas is a possible energy source for the rebrightening observed.
\item Early superhumps in EZ Lyn show roughly the same color behavior as those of V455 And and OT J012059.6+325545.

\item Ordinary superhumps in EZ Lyn show a clear anticorrelation between color and brightness, in contrast to that seen in V455 And.
This difference was possibly caused by the difference in the disk size when the ordinary superhump was excited, although the possibility of other system parameters such as orbital inclination, mass ratio, or orbital period are not excluded.
To study the cause of this diversity would require a greater number of samples of high accuracy multi-color observation.
\end{itemize}

We acknowledge with thanks the variable star observations from the AAVSO International Database contributed by observers worldwide and used in this research.
This work is supported by the MEXT-Supported Program for the Strategic Research Foundation at Private Universities (2008-2012), MEXT under Grant-in-Aid for Scientific Research (C) (22540257), and by JSPS KAKENHI Grant Number 22540252 and 25120007.
Part of this work was carried out on the open-use data analysis system at the Astronomy Data Center of the National Astronomical Observatory of Japan.

\newpage


\begin{table}
  \caption{Log and results of our photometric observations.}\label{tab:photobslog}
  \begin{center}
    \begin{tabular}{cccrccccrc}
      \hline
     \multicolumn{2}{c}{$\rm{BJD}_{\rm TDB}$($2455000+$)}  & yy/mm/dd  & T  &  $g'$  &  err  &   $i'$  &  err  &  $g'-i'$ & err  \\
     start     &    end    &          &    &        &       &        &       &        &       \\
     458.25325 & 458.31473 & 10/09/18 &  0 & 11.981 & 0.002 & 12.346 & 0.004 & $-0.365$ & 0.004 \\
     465.18840 & 465.30757 & 10/09/25 &  7 & 12.602 & 0.001 & 12.939 & 0.002 & $-0.337$ & 0.003 \\
     468.16399 & 468.33403 & 10/09/28 & 10 & 13.051 & 0.002 & 13.286 & 0.003 & $-0.235$ & 0.004 \\
     470.23177 & 470.25576 & 10/09/30 & 12 & 13.387 & 0.002 & 13.585 & 0.004 & $-0.198$ & 0.005 \\
     471.14787 & 471.27293 & 10/10/01 & 13 & 13.699 & 0.002 & 13.883 & 0.004 & $-0.184$ & 0.005 \\
     474.16463 & 474.20725 & 10/10/04 & 16 & 16.341 & 0.026 & 15.926 & 0.021 & $ 0.415$ & 0.034 \\
     476.18580 & 476.25985 & 10/10/06 & 18 & 13.855 & 0.003 & 14.127 & 0.007 & $-0.272$ & 0.008 \\
     477.14163 & 477.29174 & 10/10/07 & 19 & 15.057 & 0.006 & 15.064 & 0.009 & $-0.007$ & 0.011 \\
     480.23870 & 480.25156 & 10/10/10 & 22 & 15.085 & 0.005 & 15.149 & 0.011 & $-0.064$ & 0.012 \\
     482.25663 & 482.32014 & 10/10/12 & 24 & 13.880 & 0.004 & 14.111 & 0.007 & $-0.231$ & 0.008 \\
     485.25574 & 485.27094 & 10/10/15 & 27 & 14.539 & 0.003 & 14.786 & 0.007 & $-0.247$ & 0.008 \\
     486.30706 & 486.31550 & 10/10/16 & 28 & 15.880 & 0.008 & 15.625 & 0.013 & $ 0.255$ & 0.015 \\ 
     502.08360 & 502.14949 & 10/11/01 & 44 & 17.148 & 0.007 & 16.833 & 0.007 & $ 0.315$ & 0.010 \\
     507.10757 & 507.17581 & 10/11/06 & 49 & 17.366 & 0.009 & 17.255 & 0.012 & $ 0.111$ & 0.015 \\
     532.32040 & 532.36975 & 10/12/01 & 74 & 17.509 & 0.009 & 17.744 & 0.017 & $-0.235$ & 0.019 \\
     569.27492 & 569.28571 & 11/01/07 &111 & 17.564 & 0.009 & 17.818 & 0.017 & $-0.254$ & 0.019 \\
      \hline
    \end{tabular}
  \end{center}
\end{table}

\begin{table}
  \caption{Log of our spectroscopic observation}\label{tab:specobslog}
  \begin{center}
    \begin{tabular}{ccclccc}
      \hline
     \multicolumn{2}{c}{$\rm{BJD}_{\rm TDB}$($2455000+$)}  & yy/mm/dd  & T${}_\mathrm{exp}$[s] x N  & orbital phase  &  $\Delta\rm{BJD}/P_{\rm orb}$\footnotemark[1] &  superhump phase \\
     start     &    end    &           &          &             &       &              \\
     465.31932 & 465.34031 &  10 09 25 & 300 x  6 &  0.20--0.55  &  0.36 &   0.15--0.50  \\
     471.28038 & 471.30432 &  10 10 01 & 300 x  6 &  0.22--0.63  &  0.41 &   0.15--0.55  \\
     477.30235 & 477.33754 &  10 10 07 & 300 x 10 &  0.28--0.88  &  0.60 &     --       \\
     482.32535 & 482.33933 &  10 10 12 & 300 x  4 &  0.41--0.65  &  0.24 &     --       \\
     483.14357 & 483.16105 &  10 10 13 & 300 x  5 &  0.28--0.58  &  0.30 &     --       \\
     486.26559 & 486.29362 &  10 10 16 & 600 x  4 &  0.19--0.67  &  0.48 &     --       \\
      \hline
    \end{tabular}
  \end{center}
     ${}^1$ the ratio of the duration of observation ($\Delta\rm{BJD} \equiv \rm{BJD}_{\rm TDB}({\rm end}) - \rm{BJD}_{\rm TDB}({\rm start}) $) to the orbital period $P_{\rm orb}$.\\
\end{table}

\begin{table}
  \caption{Results of two-Gaussian fits for H$\alpha$ line.}\label{tab:spec_Ha_gaussian_fits}
  \begin{center}
    \begin{tabular}{cccccccc}
      \hline
               &        \multicolumn{3}{c}{Blue component}                 &        \multicolumn{3}{c}{Red component}                  &  Red $-$ Blue   \\
             T & $\lambda_\mathrm{c}$ &       EW         &      FWHM       & $\lambda_\mathrm{c}$ &       EW         &       FWHM      & $\Delta\lambda_\mathrm{c}$\\
               &        [nm]          &      [nm]        &       [nm]      &        [nm]          &      [nm]        &       [nm]      &      [nm]       \\      
             7 &  655.46 $\pm$ 0.19   & \phantom{$-$}0.04 $\pm$ 0.02 & 1.08 $\pm$ 0.37 &  657.11 $\pm$ 0.09   & $-$0.18 $\pm$ 0.02 & 1.74 $\pm$ 0.19 &       \\ 
            13 &  654.86 $\pm$ 0.11   & $-$0.26 $\pm$ 0.04 & 1.90 $\pm$ 0.23 &  657.33 $\pm$ 0.08   & $-$0.35 $\pm$ 0.04 & 1.93 $\pm$ 0.18 & 2.46 $\pm$ 0.14 \\ 
            19 &  655.26 $\pm$ 0.06   & $-$0.82 $\pm$ 0.04 & 2.70 $\pm$ 0.13 &  657.68 $\pm$ 0.07   & $-$0.72 $\pm$ 0.05 & 2.80 $\pm$ 0.17 & 2.43 $\pm$ 0.09 \\ 
            24 &  655.15 $\pm$ 0.12   & $-$0.31 $\pm$ 0.04 & 2.45 $\pm$ 0.27 &  657.37 $\pm$ 0.11   & $-$0.37 $\pm$ 0.04 & 2.45 $\pm$ 0.23 & 2.22 $\pm$ 0.16 \\ 
            25 &  655.55 $\pm$ 0.16   & $-$0.92 $\pm$ 0.16 & 2.45 $\pm$ 0.36 &  657.18 $\pm$ 0.30   & $-$0.63 $\pm$ 0.20 & 3.20 $\pm$ 0.88 & 1.63 $\pm$ 0.34 \\ 
            28 &  654.47 $\pm$ 0.36   & $-$0.73 $\pm$ 0.27 & 2.88 $\pm$ 0.94 &  657.25 $\pm$ 0.14   & $-$0.86 $\pm$ 0.22 & 1.89 $\pm$ 0.36 &                 \\ 
    quiescence\footnotemark[1] &  655.00 $\pm$ 0.01   & $-$4.98 $\pm$ 0.05 & 1.80 $\pm$ 0.02 &  657.70 $\pm$ 0.01   & $-$4.75 $\pm$ 0.05 & 1.72 $\pm$ 0.02 & 2.71 $\pm$ 0.01 \\ 
      \hline
    \end{tabular}
  \end{center}
  ${}^1$ spectrum in quiescence was obtained from SDSS DR8 archive (spec-1780-53090-0431.fits, observed on March 26, 2004). \\
\end{table}

\begin{table}
  \caption{System properties of V455 And and EZ Lyn}\label{tab:system_info_V455And_vs_EZLyn}
  \begin{center}
    \begin{tabular}{cccc}
      \hline
     Name    & P${}_\mathrm{orb}$ [d] & $\epsilon$ & $q$\footnotemark[1] \\
     V455 And & 0.05631\footnotemark[2] & 0.015\footnotemark[3] & 0.074 \\
     EZ Lyn   & 0.05900\footnotemark[4] & 0.011\footnotemark[5] & 0.056 \\
      \hline
    \end{tabular}\\
  ${}^1$ estimated by the relation between $\epsilon$ and $q$ in \citet{patterson+2005}.\\
  ${}^2$ \citet{araujo+2005} \\
  ${}^3$ \citet{kato+2009a} \\
  ${}^4$ \citet{kato+2009b} \\
  ${}^5$ \citet{kato+2012} \\
  \end{center}
\end{table}

\bigskip

\end{document}